\documentclass{amsart}

\usepackage[latin1]{inputenc}
\usepackage[english]{babel}
\usepackage[T1]{fontenc}
\usepackage{amsfonts, amsmath, amssymb, bbm}
\usepackage{amsthm}

\usepackage{graphicx}

\newtheorem{theorem}{Theorem}[section]

\theoremstyle{definition}
\newtheorem{definition}[theorem]{Definition}
\newtheorem{example}[theorem]{Example}
\newtheorem{paradigm}[theorem]{Paradigm}

\newtheorem{remark}[theorem]{Remark}

\numberwithin{equation}{section}

\newcommand{\bib}[4]{#1 (#2). #3. \textit{#4}.}
\newcommand{\cds}{\ensuremath{\mathrm{CDS}}}
\newcommand{\dom}{\ensuremath{\mathrm{dom}}}
\renewcommand{\for}{\ensuremath{\mathrm{for}}}
\newcommand{\F}{\mathbb{F}}
\renewcommand{\i}{\ensuremath{\mathbbm{1}}}
\newcommand{\integ}[4]{\ensuremath{\int_{#1}^{#2}#3\,d#4}}
\newcommand{\leli}{\ensuremath{\textrm{--}}}

\newcommand{\M}[1]{\mathcal{#1}}
\newcommand{\N}{\mathbb{N}}
\newcommand{\pderiv}[1]{\frac{\partial}{\partial#1}}
\renewcommand{\P}{\mathbb{P}}
\newcommand{\pd}{\ensuremath{\mathrm{PD}}}
\newcommand{\Q}{\mathbb{Q}}
\newcommand{\R}{\mathbb{R}}
\newcommand{\set}[2]{\ensuremath{\big\{#1\,\big|\,#2\big\}}}
\newcommand{\Set}[2]{\ensuremath{\Big\{#1\,\Big|\,#2\Big\}}}
\renewcommand{\tilde}[1]{\widetilde{#1}}

\allowdisplaybreaks

\begin{document}

\hyphenpenalty=10000

\title{The Jarrow \& Turnbull Setting Revisited}


\author{Thomas Krabichler}
\address{}
\curraddr{}
\email{thomas.krabichler@ost.ch}
\thanks{}

\author{Josef Teichmann}
\address{}
\curraddr{}
\email{josef.teichmann@math.ethz.ch}
\thanks{}


\date{}

\begin{abstract}
We consider a financial market with zero-coupon bonds that are exposed to credit and liquidity risk. We revisit the famous Jarrow \& Turnbull setting \cite{turnbull} in order to account for these two intricately intertwined risk types. We utilise the foreign exchange analogy that interprets defaultable zero-coupon bonds as a conversion of non-defaultable foreign counterparts. The relevant exchange rate is only partially observable in the market filtration, which leads us naturally to an application of the concept of \emph{platonic financial markets} as introduced in \cite{platonic}. We provide an example of tractable term structure models that are driven by a two-dimensional affine jump diffusion. Furthermore, we derive explicit valuation formulae for marketable products, e.g., for credit default swaps.
\end{abstract}

\maketitle


\section{Introduction}
A zero-coupon bond is a financial contract that promises its holder the payment of one monetary unit at maturity, with no intermediate payments. If adverse circumstances occur over the lifetime of the contract, the final redemption may only be a fraction of the promised payoff. The uncertainty about the recoverability of financial entitlements is referred to as \emph{default risk}. Even though public awareness is often not
raised sufficiently, most financial contracts that one encounters in the real world are subject to default risk. The distinction between defaultable and non-defaultable zero-coupon bonds is mainly associated with aspects of the final payoff. Even if an issuer of defaultable zero-coupon bonds manages to meet their final obligations in the end, losses could occur for some investors during the lifetime of the contract all the same. This is particularly the case, if investors do not intend to keep the contract until maturity. Rumours about the credit quality of the issuer or a respective down-grading by an external rating agency may affect the resale value adversely. Possible losses prior to maturity for the aforesaid reasons are usually referred to as \emph{migration risk}. Default risk and migration risk are generally subsumed under the broad term \emph{credit risk}.

Indisputably, there is an intricate connection between credit risk and aspects of \emph{liquidity}. These are diverse and relate to both the market and the commodity itself. More precisely, they are the \emph{asset liquidity}, describing the immediacy and the transaction cost with which the asset in scope can be converted into legal tender, and the \emph{institutional liquidity}, standing for the considered issuer's ability to meet its settlement obligations; see also the IMF Working Paper WP/02/232 \cite{imf}. Credit and liquidity risk have received a lot of attention, especially since the subprime crisis struck the financial markets in 2007/2008.

It is the aim of this article to present a neat mathematical framework that captures both phenomena. To this end, we utilise the foreign exchange (FX) analogy for credit risk modelling, which was introduced by Jarrow \& Turnbull in 1991; see \cite{turnbull}. They assumed inherently that, if a default has occurred, the recovery rate is known instantaneously. In the light of typically observing long and complicated unwinding processes, we allow ourselves to slightly modify their setting. We propose that all one can observe in the market filtration is the occurrence of a \emph{liquidity squeeze}, which results in a delay of due payments. The final recovery will only be known after a while. If it happens to be strictly lower than one monetary unit, the liquidity squeeze has turned into a \emph{default event}. This premise naturally motivates the existence of two filtrations. There is the filtration of full information on the one hand and that of genuinely observable market information on the other hand. We formalise this idea by applying a recently found fundamental theorem, see \cite{platonic}, while maintaining a tractable and arbitrage-free framework as demonstrated by our example.

Roughly speaking, there are two common concepts to study credit risk, namely \emph{structural} and \emph{reduced-form} approaches. Either of them may be understood in the FX-analogy. Though, the design, the perspective and the assumptions are different. It is worth noticing that the idea of the FX-analogy fell into oblivion as quickly as it appeared on the scene and several conceptual problems remained unsolved. Essentially, it was not debated how to deal with unobservable quantities such as, for instance, the FX rate. Moreover, clear notions for aspects of liquidity were not defined. Bianchetti revived in 2009 the idea of the FX-analogy in order to explain the occurrence of \emph{multiple yield curves}; see \cite{bianchetti}. Jarrow \& Turnbull studied the FX-analogy originally in a discrete time tree model as well as in a simple HJM-setup. They extended and refined it subsequently in \cite{jlt}, but moved on to a rather intensity-based approach for rating purposes. Up until the previously mentioned financial crisis, the FX-analogy was cited only sporadically. Nguyen \& Seifried elaborate in \cite{nguyen} on the FX-analogy when they apply the potential approach from \cite{rogers} in order to model multiple yield curves. The FX-analogy is also outlined briefly in the appendix of \cite{christa} or in Section~3.3 of \cite{grbac}, where the post-crisis LIBOR market is modelled in both cases in terms of a multicurrency HJM-framework. The FX-analogy can as well be applied for modelling inflation-linked risks; e.g., see Section~15.1 in \cite{brigo} or \cite{yildirim}.

The structure of the article is as follows. In Section~\ref{sec:jt}, we recall the essential elements of the FX-like venture. Subsequently, we specify the credit and liquidity risk terminology in Section~\ref{sec:tacs}. Next, we move on to briefly discuss topics of the FX-like venture that matter from a practical viewpoint. Finally, we present the centrepiece of this article in Section~\ref{sec:porr}. In the last section, we derive explicit valuation formulae for marketable products, e.g., for credit default swaps.

\section{The Jarrow \& Turnbull Setting}\label{sec:jt}

Let $(\Omega,\mathcal{G},\P)$ with $\mathbb{G}=(\M{G}_t)_{t\geq 0}$ be a filtered probability space satisfying the usual conditions. We consider $\P$ as objective probability measure. By $B=(B_t)_{t\geq 0}$ we describe the accumulation of the domestic risk-free bank account with initial value of one monetary unit. For any $T\geq0$, we denote by $\big(P(t,T)\big)_{0\leq t\leq T}$ the c\`adl\`ag price process of a non-defaultable zero-coupon bond with maturity $T\geq 0$ and payoff $P(T,T)=1$. Furthermore, we denote by $\big(\tilde{P}(t,T)\big)_{0\leq t\leq T}$ the c\`adl\`ag price process of a defaultable zero-coupon bond with the same maturity and a random payoff $0<\tilde{P}(T,T)\leq 1$. We assume that $P(T,T)$ and $\tilde{P}(T,T)$ are written in the same currency. The distribution of the final recovery $\tilde{P}(T,T)$ is strongly linked to the riskiness of the issuer's business model. The mappings $\omega\longmapsto P\big(t,T\big)(\omega)$ and $\omega\longmapsto \tilde{P}\big(t,T\big)(\omega)$ ought to be positive and $\M{G}_t$-measurable for all $0\leq t\leq T<\infty$. We may introduce another term structure $\big\{Q(t,T)\big\}_{0\leq t\leq T<\infty}$ via
\[Q(t,T):=\frac{\tilde{P}(t,T)}{\tilde{P}(t,t)}.\]
Note that we have $Q(T,T)=1$ and, hence, that this synthetic series is default-free. By setting $S_t:=\tilde{P}(t,t)$, we get
\begin{equation}\label{eq:jt}\tilde{P}(t,T)=S_tQ(t,T).\end{equation}
Although this rewriting is very elementary, it opens an extremely nice modelling opportunity for defaultable zero-coupon bonds. We recognise that credit risk can be analysed in an FX-like setting.

\begin{paradigm}[Jarrow \& Turnbull 1991]\label{par:jt91}
The series $P(t,T)$ and $Q(t,T)$ are considered as non-defaultable zero-coupon bonds in different currencies. $\tilde{P}(t,T)$ may be interpreted as conversion of foreign default-free counterparts. $S_t=\tilde{P}(t,t)$ is referred to as \emph{recovery rate} or \emph{spot FX rate}.
\end{paradigm}

The foreign market describes the unique default-free interest rate model in which yields are driven by $\big\{\tilde{P}(t,T)\big\}_{0\leq t\leq T<\infty}$ and obligations are always met. In multi-currency settings, $\big\{Q(t,T)\big\}_{0\leq t\leq T<\infty}$ is the main driver for so-called quanto securities denominated in the domestic currency. Their basic feature is that they are not exposed to any FX-risks whatsoever for the buyer, but certainly for the issuer. The involved payoffs are constituted as if the FX rate were kept constant after conclusion of the deal.

We shall deal with different informational structures here: it is natural to assume that the recovery rate $S_t$, or spot FX rate in our analogy, is not observable by the trader's filtration $\mathbb{F}$ at time $t$.

\begin{paradigm}\label{par:platonic}
We denote the trader's filtration by $\mathbb{F}\subset\mathbb{G} $ and we assume that the bond prices of the domestic market $P(.,T)$ are $\mathbb{F}$-adapted for $T\geq 0$, but $\tilde{P}(.,T)$ and $S$ are not necessarily.
\end{paradigm}

For such a two-filtration setting, the findings of \cite{platonic} can be applied: in order to guarantee absence of arbitrage we therefore assume that existence of a measure $\Q\approx\P$ with respect to the $\mathbb{F}$-adapted bank account num\'eraire $B$ such that the optionally projected discounted processes
\[\frac{P(t,T)}{B_t}= E_\Q\bigg[\frac{P(t,T)}{B_t}\bigg|\M{F}_t\bigg],\qquad E_\Q\bigg[\frac{S_tQ(t,T)}{B_t}\bigg|\M{F}_t\bigg]=E_\Q\bigg[\frac{\tilde{P}(t,T)}{B_t}\bigg|\M{F}_t\bigg]\]
for $0\leq t\leq T$ and each $T\geq 0$ form $\Q$-martingales.

In particular, if $\F=\mathbb{G}$, it holds that
\begin{equation}\label{eq:fwdfx}\tilde{P}(t,T)=S_t\frac{Q(t,T)}{P(t,T)}P(t,T)=E_{\Q^T}\big[S_T\big|\M{F}_t\big]P(t,T)\end{equation}
for all $0\leq t\leq T<\infty$, where
\[\frac{d\Q^T}{d\Q}\bigg|_{\M{F}_t}:=\frac{P(t,T)}{P(0,T)B_t}\]
denotes the domestic $T$-forward measure. In multi-currency settings, the quantity
\begin{equation}\label{eq:ft}F(t,T):=\frac{\tilde{P}(t,T)}{P(t,T)}=S_t\frac{Q(t,T)}{P(t,T)}\end{equation}
is usually referred to as \emph{forward FX rate}, i.e., as seen from time $t$, the agreement to exchange one foreign monetary unit for locked-in $F(t,T)$ domestic monetary units at time $T$ is at arm's length and worth zero. Thus, $F(t,T)$ is a natural basis for currency forwards. It is the risk-neutral $t$-forecast of the recovery/FX rate at time $T$. In the following, we shall use the term \emph{forward recovery rate} for $F(t,T)$ equivalently.

If the term structures $T\longmapsto P(t,T)$ and $T\longmapsto Q(t,T)$ are assumed to be $C^1$ as in the classical HJM-framework, one may consider the continuously compounded instantaneous forward rates
\[f_\dom(t,T):=-\pderiv{T}\log{P(t,T)},\qquad f_\for(t,T):=-\pderiv{T}\log{Q(t,T)}.\]
On the one hand, \eqref{eq:jt} and the product rule for logarithms yield
\begin{align*}
-\pderiv{T}\log{\tilde{P}(t,T)}&=-\pderiv{T}\log{S_t}-\pderiv{T}\log{Q(t,T)}=f_\for(t,T).
\intertext{On the other hand, utilising \eqref{eq:ft} gives}
-\pderiv{T}\log{\tilde{P}(t,T)}&=-\pderiv{T}\log{P(t,T)}-\pderiv{T}\log{F(t,T)}\\
&=f_\dom(t,T)-\pderiv{T}\log{F(t,T)}.
\end{align*}
Thus, the negative logarithmic derivative of the forward recovery rate describes the spread of the foreign forward rates above their domestic counterparts. High spreads indicate a precarious period of financial distress. In this sense, $-\pderiv{T}\log{F(t,T)}$ may be attributed to the intensity of the credit risk.

\begin{remark}[Construction of Credit Risk Models]
The FX-like approach can be implemented easily by considering a genuine multi-currency setting and restricting oneself to a certain class of FX rates, which take values only in the target zone $(0,1]$. Equation~\eqref{eq:fwdfx} is very helpful in this regard. Exemplarily, inspired by Dirichlet problems related to the Beta distribution on $(0,1)$ (e.g., see Table 1 in \cite{dirichlet}) and FX rate models in a target zone (e.g., see \cite{targetzone}), one might model the recovery rate $S=(S_t)_{t\geq 0}$ under the pricing measure $\Q$ as a Jacobi process. Alternatively, one might consider a recovery rate $S_t=e^{-\langle\xi,X_t\rangle}$ for a $d$-dimensional non-negative affine Markov process $X=(X_t)_{t\geq 0}$ and a parameter $\xi\in\R_+^d$. Both assumptions lead to fairly tractable models. This prevails also in the presence of jumps; see \cite{krabi}.
\end{remark}

\section{Credit and Liquidity Risk Terminology}\label{sec:tacs}

For the Jarrow \& Turnbull setting, we introduce the following terminology.

\begin{definition}[Defaultable Term Structures]
A zero-coupon bond with price process $\big(\tilde{P}(t,T)\big)_{0\leq t\leq T}$ for a fixed maturity $T>0$ is called \emph{defaultable}, if it holds $\P\big[S_T<1\big]>0$; otherwise it is called \emph{default-free} or \emph{non-defaultable}. A whole term structure of zero-coupon bonds $\big\{\tilde{P}(t,T)\big\}_{0\leq t\leq T<\infty}$ is called \emph{defaultable}, if for any bound $T^*>0$ there exists another maturity $T'>T^*$ for which $\big(\tilde{P}(t,T')\big)_{0\leq t\leq T'}$ is defaultable.
\end{definition}

In order not to end up in a standard framework, we will tacitly assume for the remainder of this article that $\big\{\tilde{P}(t,T)\big\}_{0\leq t\leq T<\infty}$ is defaultable. Generally, the notion \emph{riskiness} for a zero-coupon bond or a whole term structure does not relate to the uncertainty about the underlying payoffs, but to the stochastic nature of its price evolution. Still, we use the terms \emph{risk-free} for non-defaultable and \emph{risky} for defaultable interchangeably. Loans are usually only marketable for virtuous issuers who are duly concerned about their obligations.

\begin{definition}[Proper Term Structures]
A defaultable zero-coupon bond with price process $\big(\tilde{P}(t,T)\big)_{0\leq t\leq T}$ for a fixed maturity $T>0$ is called \emph{proper}, if it holds $\P\big[S_T=1\big]>0$; otherwise it is called \emph{improper}. A whole term structure of zero-coupon bonds $\big\{\tilde{P}(t,T)\big\}_{0\leq t\leq T<\infty}$ is called \emph{proper}, if any zero-coupon bond $\big(\tilde{P}(t,T)\big)_{0\leq t\leq T}$ for $T>0$ is proper.
\end{definition}

Any downturn of the recovery rate relates to a shortage of liquid funds.

\begin{definition}[Liquidity Squeeze]
Let the term structure $\big\{\tilde{P}(t,T)\big\}_{0\leq t\leq T<\infty}$ be risky. A \emph{liquidity squeeze} at time $t$ is the probabilistic event $\{S_t<1\}$.
\end{definition}

Nonetheless, a liquidity squeeze may be without consequences as long as no physical payments become due. The instances in which payments have to be settled are described by means of a sequence of stopping times. The first time a liquidity squeeze coincides with a payment date, a default event is deemed to occur.

\begin{definition}[Payment Schedule and Default Event]\label{def:defaultevent}
Let $\big\{\tilde{P}(t,T)\big\}_{0\leq t\leq T<\infty}$ be risky. Denote by $(\tau_n)_{n\in\N}$ a \emph{payment schedule}, that is a sequence of finite $\F$-stopping times without accumulation points. The \emph{default time} $\tau$ is defined as the possibly infinite stopping time $\tau:=\inf\set{\tau_n}{S_{\tau_n}<1,n\in\N}$. A \emph{default event} at time $t$ is the probabilistic event $\{\tau=t\}\cap\{S_t<1\}$.
\end{definition}

The default time $\tau$ really is a stopping time, since we have
\[\{\tau\leq t\}=\bigcup_{n\in\N}\Big(\{\tau_n\leq t\}\cap\{S_{\tau_n\wedge t}<1\}\Big).\]
Typically, a default event triggers either a restructuring or a liquidation of the issuer. As the case may be, the issuer of the defaultable bonds succeeds in reviving the business model in the aftermath of a default event. Because $S_\tau$ might not be readily observable in a sub-filtration describing genuinely accessible market data, the event $\{S_\tau=1\}$ is not unlikely; see also the discussion in the next section. Consequently, one can refine the modelling approach by allowing for multiple defaults. Notably, the occurrence of $\{S_\tau<1\}$ does not prevent the recovery rate from returning to the level $1$.

\begin{remark}[Bankruptcy]
\textup{
It is tempting to postulate that the occurrence of $\{S_t<1\}$ always causes bankruptcy of the referenced entity. However, the FX-like setting can also be associated with an arbitrary investment portfolio in defaultable corporate bonds. In that case, the resulting recovery rate is a superposition of many defaultable bonds and respective recoveries. Thus, full recovery may almost never be given and $\{S_t<1\}$ may occur without further consequences for the portfolio's existence.
}\hfill$\Box$
\end{remark}

\section{Practical Matters}

Realistically or symptomatic of general market models, one realisation of the stochastic processes $\big(P(t,T_i)\big)_{0\leq t\leq T_i}$ for $i=0,1,2,\hdots n$ and $\big(\tilde{P}(t,\tilde{T}_j)\big)_{0\leq t\leq\tilde{T}_j}$ for $j=0,1,2,\hdots\tilde{n}$ can be inferred from market quotes for two generally differing grids of rather remote maturities $0\leq T_0<T_1<T_2<\hdots<T_n$ together with $0\leq\tilde{T}_0<\tilde{T}_1<\tilde{T}_2<\hdots<\tilde{T}_{\tilde{n}}$. Hence, neither $S_t=\tilde{P}(t,t)$ nor $\underset{T\to t+}{\lim}\tilde{P}(t,T)$ are always accessible, given the latter expression makes sense at all. In contrast, the short end $P(t,t)=1$ of the non-defaultable term structure is known at any time $t\geq 0$. Thus, the extrapolation exercise is much easier for the non-defaultable term structure than it is for its defaultable counterpart; yet the non-defaultable short rate
\[r_t=-\pderiv{T}\bigg|_{T=t}\log{P(t,T)},\]
given this notion makes sense at all, is normally also unknown. The third series of zero-coupon bonds $\big\{Q(t,T)\big\}_{0\leq t\leq T<\infty}$ is synthetic and its introduction is subject to a redundancy. As $S_t$ and $Q(t,T)/P(t,T)$ are only observable sporadically and predominantly only on aggregated level $F(t,T)$, the market (more precisely, a certain sub-filtration of $\mathbb{G}$ describing the genuinely observable events) captures only very little information of the FX-like setting. The recovery rate is only observable on a discrete payment schedule; see Definition~\ref{def:defaultevent} above. Therefore, further model assumptions have to be imposed. This is in strong contrast to multi-currency settings in which the spot FX rate is usually known. However, this is not of utmost relevance in some applications. Typically, one assumes full recovery anyway and is rather interested in the likelihood of future downturns. Given full recovery $\{S_t=1\}$, marketable instruments and credit derivatives can help to infer spread factors $Q(t,T)/P(t,T)$ and to calibrate the models against the current market conditions.

Mathematically, the situation can be modelled as described in Paradigm~\ref{par:platonic}. One equips the general FX-like setting with two filtrations. The trader's filtration $\F$ comprises only genuinely observable market information and a larger filtration $\mathbb{G}$ also carries hidden information; the recovery rate process is $\mathbb{G}$-adapted but not necessarily $\F$-adapted. Both filtrations are augmented in order to satisfy the usual conditions. Let $(\tau_n)_{n\in\N}$ denote a payment schedule, $\tau$ the $[0,\infty]$-valued default time and $\tau'\geq\tau$ be an $\F$-stopping time that fixes the final recovery for general market participants. Besides the price evolution of selected bonds up to time $t$, the core part of the $\sigma$-algebra $\M{F}_t$ is generated by the sets $\set{\{\tau_n\leq u\}}{u\leq t}$, which simply are the payment dates,
\begin{align*}
&\Set{\{\tau_n\leq u\}\cap\big\{S_{\tau_n}\in A\big\}}{n\in\N,u\leq t,A\in\big\{(0,1),\{1\}\big\}},
\intertext{which distinguish payment dates qualitatively on whether any liquidity squeeze has occurred previously or not, and, for the Borel $\sigma$-algebra $\M{B}$,}
&\Set{\{\tau'\leq u\}\cap\big\{S_\tau\in A\big\}}{u\leq t,A\in\M{B}\big((0,1]\big)},
\end{align*}
which is the knowledge about the final post-default recovery. The filtering problem consists then in calculating $E_\P\big[S_t\big|\M{F}_t\big]$ for $t\geq 0$. It does not necessarily hold $E_\P\big[S_t\big|\M{F}_t\big]\equiv1$ on $\{\tau>t\}$, since low prices for defaultable bonds may announce an upcoming default. We are making this notion more precise in the next section.

\section{Partial Observability of the Recovery Rate}\label{sec:porr}

It is an idiosyncrasy of the general FX-like setting that, given a default has happened, the involved recovery rate is known instantaneously. From the practical viewpoint, this feature may legitimately be questioned. All one usually knows is that some sort of default event has happened. The final recovery will typically only be determined after a long and complicated unwinding process. The settlement of a default event always comes with a negotiation process. It is the aim of this section to generalise the FX-analogy in this direction. To this end, we trade off phenomenological richness against full analytical or at least numerical tractability.\\
Let us consider a two-dimensional $[0,\infty)^2$-valued conservative regular affine jump diffusion $(X,Y)$ with
\[dX_t=\sigma_X\sqrt{X_t}\,dW_t^X+dJ_t^X,\qquad dY_t=\sigma_Y\sqrt{Y_t}\,dW_t^Y+dJ_t^Y\]
and some initial condition $(X_0,Y_0)=(0,y_0)$ with $y_0\geq 0$. $(W^X,W^Y)$ is a two-dimensional Brownian motion on a filtered probability space $(\Omega,\M{F},\Q)$ with $\F=(\M{F}_t)_{t\geq 0}$ satisfying the usual conditions. $(J^X,J^Y)$ is a right-continuous pure jump process, whose jump heights have a fixed distribution $\nu$ on $(0,\infty)^2$ and arrive with intensity $m+\mu_XX_{t\leli}+\mu_YY_{t\leli}$ for some parameters $m>0$ and $\mu_X,\mu_Y\geq 0$. The Brownian motions, the jumps heights and the arrival of the jumps are assumed to be independent under the risk-neutral measure $\Q$.

\begin{figure}[htp]
\begin{center}
\includegraphics[width=\textwidth]{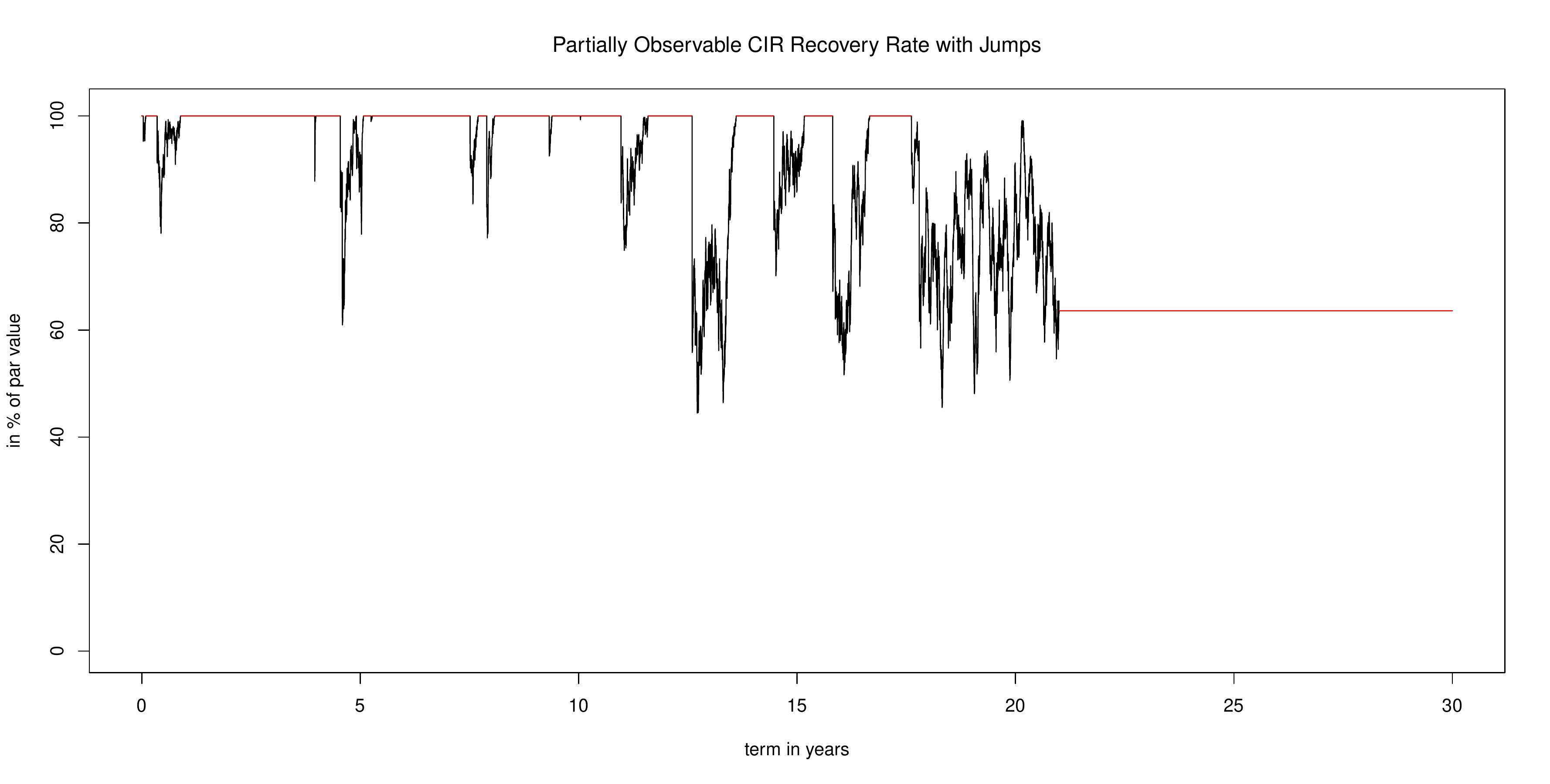}
\end{center}
\caption{The sample path illustrates the setting of this section. If liquidity squeezes prevail for too long, they turn into default events. The final recovery will only be known after a while. Furthermore, only the segments marked red are observable for general market participants. Either full institutional liquidity is given or there are payment delays. In the latter case, the actual level of the recovery rate is completely unknown.}\label{fig:riccatiex}
\end{figure}

For simplicity, we assume a trivial non-defaultable term structure $B\equiv 1$. The partially observable recovery rate is derived from the auxiliary process $e^{-X_t}$ and typically starts with a constant trajectory at the level one. $Y$ features the intensity of liquidity squeezes. Once the recovery has jumped below one, the recovery rate pursuits an unsteady course. The further $X$ jumps away from zero, the more likely become positive jumps for $Y$. This itself triggers yet another surge for a potential depreciation of the recovery rate. To this extent, downturns of the recovery rate are \emph{self-exciting}. Nonetheless, $X$ may also return to zero, since it is after all a Feller diffusion, and full recovery prevails for a certain period. See Figure~\ref{fig:riccatiex} for an illustration of the above setting. In more involved settings, $Y$ could stand for the risk-free short rate process. In this sense, defaults would be more likely with an increasing interest rate burden; see Remark~\ref{rmk:bs} below.

We assume here that the trader's filtration $\F$ is generated by $ X,Y $. This means if a liquidity squeeze occurs and contingencies of a zero-coupon bond maturing at time $T$ cannot be paid off, this is usually known. However, the actual recovery is generally not observable: we assume that $\tilde P(T,T)$ pays off
\[\i_{\{X_T=0\}}+e^{-X_{T+h}}\i_{\{X_T>0\}}\]
at the time instances $T$ or $T+h$ respectively, where $h>0$ is a positive parameter. Whence, this corresponds to $\M{G}_t=\M{F}_{t+h}$ in the language of the two-filtration setting. By utilising the affine Markov structure, we aim at deriving an explicit valuation formula for a defaultable zero-coupon bond $\tilde{P}(t,T)$, i.e., we calculate the optional projections on $\mathbb{F}$. It is remarkable that this is possible in this setting.

Consistent with the above premises, the generalised Riccati equations\index{Riccati equations} for $u,v\in\R$ and $0\leq t\leq T$ are given by
\begin{align*}
\partial_t\phi(t,iu,iv)&=m\kappa\big(\psi_X(t,iu,iv),\psi_Y(t,iu,iv)\big),\\
\partial_t\psi_X(t,iu,iv)&=\frac{1}{2}{\sigma_X}^2{\psi_X(t,iu,iv)}^2+\mu_X\kappa\big(\psi_X(t,iu,iv),\psi_Y(t,iu,iv)\big),\\
\partial_t\psi_Y(t,iu,iv)&=\frac{1}{2}{\sigma_Y}^2{\psi_Y(t,iu,iv)}^2+\mu_Y\kappa\big(\psi_X(t,iu,iv),\psi_Y(t,iu,iv)\big)
\intertext{with the initial conditions $\phi(0,iu,iv)=0$, $\psi_X(0,iu,iv)=iu$ and $\psi_Y(0,iu,iv)=iv$, where}
\kappa(iu,iv)&:=\integ{(0,\infty)^2}{}{\big(e^{iux+ivy}-1\big)}{\nu(x,y)};
\end{align*}
see Theorem~2.7 in \cite{affgen} for a general treatment. Accordingly, it holds for all $u,v\in\R$ and $0\leq t\leq T$
\[E_\Q\Big[e^{iuX_T+ivY_T}\Big|\M{F}_t\Big]=e^{\phi(T-t,iu,iv)+X_t\psi_X(T-t,iu,iv)+Y_t\psi_Y(T-t,iu,iv)},\]
where $\M{F}_t=\sigma\big(X_u,Y_u;u\leq t\big)$. This can be exploited for the implied forward recovery rates, since
\begin{align}
F(t,T)&=E_\Q\Big[\i_{\{X_T=0\}}+e^{-X_{T+h}}\i_{\{X_T>0\}}\Big|\M{F}_t\Big]\nonumber\\
&=E_\Q\Big[\i_{\{X_T=0\}}+E_\Q\big[e^{-X_{T+h}}\big|\M{F}_T\big]\i_{\{X_T>0\}}\Big|\M{F}_t\Big].\label{eq:projtemp}
\end{align}
Due to the Riemann-Lebesgue lemma, the mass of the atom $\{X_T=0\}$ can be calculated in terms of
\[\Q\big[X_T=0\big]=\lim_{u\to\infty}E_\Q\big[e^{iuX_T}\big]=\lim_{u\to\infty}e^{\phi(T,iu,0)+y_0\psi_Y(T,iu,0)}.\]
The projection of the second summand in \eqref{eq:projtemp} can be rewritten as
\begin{align*}
&E_\Q\Big[E_\Q\big[e^{-X_{T+h}}\big|\M{F}_T\big]\i_{\{X_T>0\}}\Big|\M{F}_t\Big]\\
&\ =E_\Q\Big[e^{\phi(h,-1,0)+X_T\psi_X(h,-1,0)+Y_T\psi_Y(h,-1,0)}\big(1-\i_{\{X_T=0\}}\big)\Big|\M{F}_t\Big]\\
&\ =E_\Q\Big[e^{\phi(h,-1,0)+X_T\psi_X(h,-1,0)+Y_T\psi_Y(h,-1,0)}\Big|\M{F}_t\Big]\\
&\qquad-E_\Q\Big[e^{\phi(h,-1,0)+Y_T\psi_Y(h,-1,0)}\i_{\{X_T=0\}}\Big|\M{F}_t\Big].
\end{align*}
The minuend is yet another evaluation of the Fourier transform. The subtrahend can also be calculated explicitly by utilising dominated convergence and the relation
\[E_\Q\Big[e^{vY_T}\i_{\{X_T=0\}}\Big|\M{F}_t\Big]=\lim_{u\to-\infty}E_\Q\Big[e^{uX_T+vY_T}\Big|\M{F}_t\Big].\]
Hence, L\'evy's inversion theorem is not required. Generally, already humble model assumptions lead to Riccati equations, which are intractable analytically. Therefore, numerical approximation procedures are indispensable anyway; however, the affine structure reduces the complexity of the approximation schemes considerably. The complexity remains low if one incorporates drift terms for $X$ and $Y$ in order to increase market consistency. Let us illustrate the above framework by a couple of simple examples.

\begin{example}[Deterministic Jump Intensity, $\sigma_X\neq0$]\label{ex:case1}
\textup{
As a starting point, we disable the stochastic intensity $Y$ and consider the limiting case $m>0$, $\mu_X=0$ and $\mu_Y=0$. Regarding the jump sizes, let us choose for $\lambda_X>0$ the product measure
\[d\nu(x,y)=\lambda_Xe^{-\lambda_X x}\,dx\otimes\delta_{\{0\}}(dy),\]
where the Dirac measure for $A\in\M{B}(\R)$ is defined as
\[\delta_{\{0\}}(A)=\begin{cases}1&\textrm{, if }0\in A,\\0&\textrm{, otherwise}.\end{cases}\]
Whenever a jump occurs, it is entirely in the $x$-direction. Each jump size itself is exponentially distributed with the parameter $\lambda_X$. Since $\kappa(iu,iv)=\frac{iu}{\lambda_X-iu}$, this model choice translates into the system of Riccati equations
\begin{align*}
\dot{\phi}&=\frac{m\psi_X}{\lambda_X-\psi_X},\\
\dot{\psi_X}&=\frac{1}{2}{\sigma_X}^2{\psi_X}^2
\intertext{with the solution}
\phi(t,iu,iv)&=\frac{2m}{\lambda_X{\sigma_X}^2}\log\Bigg(\frac{\frac{\lambda_X}{iu}-1}{\lambda_X\big(\frac{1}{iu}-\frac{1}{2}{\sigma_X}^2t\big)-1}\Bigg),\\
\psi_X(t,iu,iv)&=\frac{1}{\frac{1}{iu}-\frac{1}{2}{\sigma_X}^2t}.
\end{align*}
It obviously holds
\[\phi(t,iu,iv)\stackrel{u\to\infty}{\longrightarrow}\frac{2m}{\lambda_X{\sigma_X}^2}\log\bigg(\frac{1}{\frac{1}{2}\lambda_X{\sigma_X}^2t+1}\bigg).\]
This results in the initial forward recovery rate
\[F(0,T)=\big(1-e^{\phi(h,-1,0)}\big)\Q\big[X_T=0\big]+e^{\phi(h,-1,0)}E_\Q\Big[e^{X_T\psi_X(h,-1,0)}\Big],\]
where
\begin{align*}
\Q\big[X_T=0\big]&=\bigg(\frac{1}{2}\lambda_X{\sigma_X}^2T+1\bigg)^{-\frac{2m}{\lambda_X{\sigma_X}^2}}
\intertext{and}
E_\Q\Big[e^{X_T\psi_X(h,-1,0)}\Big]&=e^{\phi\big(T,\psi_X(h,-1,0),0\big)}.
\end{align*}
Thus, analytical pricing formulas are available. For time instances $t>0$, the forward recovery rates $T\longmapsto F(t,T)$ depending on the state variable $X_t$ can be calculated analogously by utilising the Markov property.
}\hfill$\Box$
\end{example}

\begin{remark}[Time Value of Money]\label{rmk:bs}
\textup{
If we modify Example~\ref{ex:case1} in the sense that $Y$ does not only describe the jump intensity, but also the risk-free short rate, then the model remains fully tractable. All one has to do is introduce an additional structural component $Z=(Z_t)_{t\geq 0}$ with
\[Z_t:=z_0+\integ{0}{t}{Y_u}{u}.\]
In this case, $(X,Y,Z)$ denotes a three-dimensional affine jump diffusion. The calculations get a bit more cumbersome, but the derivation of a closed-form valuation formula is possible all the same; e.g., see Example~3.20 in \cite{krabi}.
}\hfill$\Box$
\end{remark}

\begin{remark}[Generalisation]
\textup{
The presented recipe in order to calculate \eqref{eq:projtemp} is purposeful for the next upcoming critical maturity. For a discrete payment schedule $0<T_1<T_2$, the payoff due at time $T_2$ depends on whether a default event occurred at time $T_1$ or not. Consistently, one might consider for $0\leq t\leq T_1<T_1+h\leq T_2$ the generalised functional
\[E_\Q\bigg[\i_{\{X_{T_1+h}=0\}}\Big(\i_{\{X_{T_2}=0\}}+e^{-X_{T_2+h}}\i_{\{X_{T_2}>0\}}\Big)+e^{-X_{T_1+h}}\i_{\{X_{T_1}>0\}}\i_{\{X_{T_1+h}>0\}}\bigg|\M{F}_t\bigg].\]
For $0\leq t\leq T_1<T_2<T_1+h$, the adaptation works analogously. Explicit pricing formulae become more complex but they are still available by cascading the above recipe. This may be useful in the market-consistent valuation of fully collateralised over-the-counter deals with daily margining. If one fails to post collateral, one has a time limit of a few days $h$, as agreed upon in the indenture, to supply the due amount. If the deficiency still prevails thereafter, the deal is closed.
}\hfill$\Box$
\end{remark}

\begin{example}[$\mu_X=0$, $\sigma_X=0$]\label{ex:case2}
\textup{
We consider the case $m>0$, $\mu_X=0$ and $\mu_Y>0$ as well as $\sigma_X=0$. Correspondingly, the stochastic intensity is indifferent with respect to the level of $X$. Once a liquidity squeeze has occurred, full recovery is not possible any more. Regarding the jump sizes, we choose the same product measure as in the previous example. In this case, the Riccati equations reduce to
\begin{align*}
\dot{\phi}&=m\frac{\psi_X}{\lambda_X-\psi_X},\\
\dot{\psi_X}&=0,\\
\dot{\psi_Y}&=\frac{1}{2}{\sigma_Y}^2{\psi_Y}^2+\mu_Y\frac{\psi_X}{\lambda_X-\psi_X}
\intertext{with the solution}
\phi(t,iu,iv)&=m\frac{iu}{\lambda_X-iu}t,\\
\psi_X(t,iu,iv)&=iu,\\
\psi_Y(t,iu,iv)&=\sqrt{\frac{2\mu_Yiu}{{\sigma_Y}^2(\lambda_X-iu)}}\tan{\Bigg(C+t\sqrt{\frac{iu\mu_Y{\sigma_Y}^2}{2(\lambda_X-iu)}}\Bigg)},
\intertext{where the integration constant, certainly depending on $iu$ and $iv$,}
C&=\arctan{\Bigg(iv\sqrt{\frac{{\sigma_Y}^2(\lambda_X-iu)}{2\mu_Yiu}}\Bigg)}
\end{align*}
is chosen such that the initial condition is met. It clearly holds $\phi(t,iu,iv)\stackrel{u\to\infty}{\longrightarrow}-mt$. Moreover, since $\tan{iu}=i\tanh{u}$ for all $u\in\R$, one simply derives
\[\psi_Y(t,iu,0)\stackrel{u\to\infty}{\longrightarrow}-\sqrt{\frac{2\mu_Y}{{\sigma_Y}^2}}\tanh{\Bigg(t\sqrt{\frac{1}{2}\mu_Y{\sigma_Y}^2}\Bigg)}.\]
All in all, we get for $F(0,T)$ the expression
\begin{align}
&\Q\big[X_T=0\big]+e^{\phi(h,-1,0)+\phi\big(T,-1,\psi_Y(h,-1,0)\big)+y_0\psi_Y\big(T,-1,\psi_Y(h,-1,0)\big)}\label{eq:schnipo}\\
&\ -e^{\phi(h,-1,0)-mT+\sqrt{-\frac{2\mu_Y}{{\sigma_Y}^2}}\tan{\bigg(\arctan{\Big(\psi_Y(h,-1,0)\sqrt{-\frac{{\sigma_Y}^2}{2\mu_Y}}\Big)}+T\sqrt{-\frac{1}{2}\mu_Y{\sigma_Y}^2}\bigg)}},\nonumber
\end{align}
where
\begin{align}\label{eq:massinzero}
\Q\big[X_T=0\big]=&\ e^{-mT-y_0\sqrt{\frac{2\mu_Y}{{\sigma_Y}^2}}\tanh{\big(T\sqrt{\frac{1}{2}\mu_Y{\sigma_Y}^2}\big)}}.
\end{align}
Consequently, the state of the initial term structure is captured by the six model parameters $h$, $\lambda_X$, $m$, $\mu_Y$, $\sigma_Y$ and $y_0$. Full analytical tractability is given as in the previous example.
}\hfill$\Box$
\end{example}

\begin{remark}[Deterministic Jump Sizes]
\textup{
It is tempting to consider the above examples in the case of deterministic jump sizes. For instance, one might want to consider a product measure
\begin{equation}\label{eq:prdctmeas} d\nu(x,y)=\delta_{\{J_X\}}(dx)\otimes\delta_{\{J_Y\}}(dy)\end{equation}
for parameters $J_X>0$ and $J_Y\geq 0$. Note that if $J_X$ was set to zero, the recovery rate could not depreciate. This choice translates into $S$ having the discrete support $\set{e^{-kJ_X}}{k\in\N_0}$. Exemplarily, \eqref{eq:schnipo} and \eqref{eq:massinzero} would still prevail, if one replaced the product measure in Example~\ref{ex:case2} by \eqref{eq:prdctmeas} for $J_Y=0$, ceteris paribus. The only difference is that the parameterisation $\phi(t,iu,iv)=m\big(e^{iuJ_X}-1\big)t$ changes from $\lambda_X$ to $J_X$.
}\hfill$\Box$
\end{remark}

\begin{remark}[$\mu_X=0$, $\sigma_X\neq 0$]
\textup{
If we let $\sigma_X\neq 0$ in Example~\ref{ex:case2}, then the derivation of the forward recovery rates is more involved. The Riccati equations read
\begin{align*}
\dot{\phi}&=m\frac{\psi_X}{\lambda_X-\psi_X},\\
\dot{\psi_X}&=\frac{1}{2}{\sigma_X}^2{\psi_X}^2,\\
\dot{\psi_Y}&=\frac{1}{2}{\sigma_Y}^2{\psi_Y}^2+\mu_Y\frac{\psi_X}{\lambda_X-\psi_X}.
\end{align*}
The solutions
\[\psi_X(t,iu,iv)=\frac{2}{\frac{2}{iu}-{\sigma_X}^2t},\quad\phi(t,iu,iv)=\frac{2m}{\lambda_X{\sigma_X}^2}\log{\bigg(\frac{2\lambda_X-2iu}{2\lambda_X-2iu-iu\lambda_X{\sigma_X}^2t}\bigg)}\]
are readily at hand. If the auxiliary function $x(t)=x(t,iu,iv)$ denotes a solution to the second order linear differential equation
\[\ddot{x}+\frac{iu\mu_Y{\sigma_Y}^2}{2\lambda_X-2iu-iu\lambda_X{\sigma_X}^2t}x=0,\qquad\frac{\dot{x}(0)}{x(0)}=-\frac{iv{\sigma_Y}^2}{2},\]
then we can characterise $\psi_Y$ in terms of
\[\psi_Y(t,iu,iv)=-\frac{2\dot{x}(t)}{{\sigma_Y}^2x(t)}.\]
However, since the matrix
\[A(t)=A(t,iu,iv)=\left(\begin{matrix}0&1\\-\frac{iu\mu_Y{\sigma_Y}^2}{2\lambda_X-2iu-iu\lambda_X{\sigma_X}^2t}&0\end{matrix}\right)\]
is non-commutative for different time parameters, there is no straightforward closed-form solution for $x(t)$ as matrix exponential.
}\hfill$\Box$
\end{remark}

\begin{remark}[Link to Classical Credit Risk Models]
\textup{
The approach of this section is a neat modification of classical credit risk models. Furthermore, it incorporates naturally the phenomenon of liquidity. On the one hand, one may perceive the recovery rate process as being arisen from a not directly observable balance sheet variable. This is the structural component of the model. On the other hand, the model features in the background an intensity process for jumps that trigger both liquidity squeezes and defaults. The existence of two filtrations is idiosyncratic for doubly-stochastic settings. Similarly, defaults are not predictable. A generalisation in this regard can be found in \cite{chen} and \cite{gehmlich}.
}\hfill$\Box$
\end{remark}

\begin{remark}[Real-World vs.\ Risk-Neutral Default Probabilities]
\textup{
Sometimes, as for instance in the current mortgage business, one encounters that the defaultable term structure is rather pronounced, whereas the real-world default probabilities seem to be marginal. We can incorporate this case of a high market price of risk into our setting by considering another measure $\P\approx\Q$. Under this real-world measure, the jump size distribution of the recovery rate remains unaffected. However, the jump intensity might be arbitrarily low; e.g., see Section~4 in \cite{duffie}. $\Q$ may be seen as the result of pricing claims under $\P$ together with a highly risk averse utility function.
}\hfill$\Box$
\end{remark}

\section{Valuation Formulae for Marketable Products}

\subsection*{Mathematical Setup}

For the non-defaultable term structure $T\longmapsto P(0,T)$, we consider the \emph{extended Cox-Ingersoll-Ross} (CIR++) short rate model
\[r_t=x_t+\varphi(t),\qquad dx_t=(b_x-\beta_x x_t)\,dt+\sigma_x\sqrt{x_t}\,dW_t^x\]
for some deterministic function $t\longmapsto\varphi(t)$, positive parameters $b_x$, $\beta_x$, $\sigma_x$ with $2b_x\geq{\sigma_x}^2$ and some initial condition $r_0\in\R$; see Section~3.9 in \cite{brigo}. Then it holds
\begin{align}
P(0,T)=&\ E_\Q\Big[e^{-\integ{0}{T}{r_u}{u}}\Big]=e^{-A(0,T)-B(0,T)r_0},\label{eq:CIRnondef}\\
\intertext{where}
A(0,T)=& -\frac{2b_x}{{\sigma_x}^2}\log\bigg\{\frac{2\lambda_x e^{(\lambda_x+\beta_x)T/2}}{(\lambda_x+\beta_x)\big(e^{\lambda_x T}-1\big)+2\lambda_x}\bigg\}\nonumber\\
\ &-\varphi(0)B(0,T)+\integ{0}{T}{\varphi(u)}{u},\nonumber\\
B(0,T)=&\ \frac{2\big(e^{\lambda_x T}-1\big)}{(\lambda_x+\beta_x)\big(e^{\lambda_x T}-1\big)+2\lambda_x},\nonumber\\
\lambda_x=&\ \sqrt{{\beta_x}^2+2{\sigma_x}^2}\nonumber.
\end{align}
Regarding the defaultable term structure $T\longmapsto\tilde{P}(0,T)$, we consider the affine jump diffusion $dX_t=\sigma_X\sqrt{X_t}\,dW_t^X+dJ_t^X$ with the initial condition $X_0=x_0$. $(W^x,W^X)$ is a two-dimensional Brownian motion and $J^X$ is a right-continuous pure jump process, whose jump heights have the fixed distribution $d\nu(x)=\lambda_Xe^{-\lambda_Xx}\,dx$ and arrive with intensity $m_X+\mu_XX_{t\leli}$. The recovery rate is modelled as $S_t:=e^{-X_t}$. By construction, it holds $E_\Q\big[e^{iuX_T}\big]=e^{\phi_X(T,iu)+x_0\psi_X(T,iu)}$ for all $T\geq 0$, where $\phi_X$ and $\psi_X$ satisfy the generalised Riccati equations
\begin{align*}
\dot{\phi}_X(t,iu)&=m_X\frac{\psi_X(t,iu)}{\lambda_X-\psi_X(t,iu)},\\
\dot{\psi}_X(t,iu)&=\frac{1}{2}{\sigma_X}^2\psi_X(t,iu)^2+\mu_X\frac{\psi_X(t,iu)}{\lambda_X-\psi_X(t,iu)}.
\end{align*}
Provided that $\mu_X=0$, these equations can be solved explicitly;
\begin{align*}
\phi_X(t,iu)&=\frac{2m_X}{\lambda_X{\sigma_X}^2}\log\Bigg(\frac{\frac{\lambda_X}{iu}-1}{\lambda_X\big(\frac{1}{iu}-\frac{1}{2}{\sigma_X}^2t\big)-1}\Bigg),\\
\psi_X(t,iu)&=\frac{1}{\frac{1}{iu}-\frac{1}{2}{\sigma_X}^2t}.
\end{align*}
In general, the expected recovery is given by
\begin{align}
F(0,T)&=\frac{\tilde{P}(0,T)}{P(0,T)}=E_\Q\big[S_T\big]=E_\Q\big[e^{-X_T}\big]=e^{\phi_X(T,-1)+x_0\psi_X(T,-1)}.\label{eq:f}
\intertext{Provided that $\mu_X=0$, it holds}
F(0,T)&=\bigg(\frac{\lambda_X+1}{\lambda_X(1+\frac{1}{2}{\sigma_X}^2T)+1}\bigg)^{\frac{2m_X}{\lambda_X{\sigma_X}^2}}e^{\frac{-x_0}{1+\frac{1}{2}{\sigma_X}^2T}}.\nonumber
\end{align}

\subsection*{Marketable Bonds}

For some parameter $n$ and maturity $T$, let $0=T_0<T_1<T_2<\hdots<T_n=T$ denote a partition of $[0,T]$. On the one hand, the initial value of a non-defaultable government bond paying $n$ coupons of annualised size $c$ at the time instances $T_1,T_2,\hdots,T_n$ is given by
\begin{equation}\label{eq:govbonds}V_{n,c}(0,T)=P(0,T)+c\sum_{i=1}^n(T_i-T_{i-1})P(0,T_i).\end{equation}
On the other hand, the initial value of a defaultable corporate bond paying $n$ coupons of annualised size $c$ at the time instances $T_1,T_2,\hdots,T_n$ is given consistently by
\begin{equation}\label{eq:corpbonds}\tilde{V}_{n,c}(0,T)=\tilde{P}(0,T)+c\sum_{i=1}^n(T_i-T_{i-1})\big(\tilde{P}(0,T_i)+\mathbb{L}(0,T_i)\big).\end{equation}
The additional term structure $T\longmapsto\mathbb{L}(0,T)$ accounts for asset illiquidity. From the mathematical viewpoint, as proposed in \cite{krabi}, the term structure $T\longmapsto\mathbb{L}(0,T)$ can be modelled by an illiquidity deflator $Z$, which is a strict $\Q$-local martingale. $Z$ is interpreted as value process of a roll-over strategy in defaultable zero-coupon bonds. The resulting discrepancy $\mathbb{L}(0,T)$ between the fundamental value and the market value of a defaultable bond cannot be exploited due to admissibility constraints. The raw spread between $P(0,T)$ and $\tilde{P}(0,T)$ can be determined with another marketable product that we tackle in the next section.

\subsection*{Credit Default Swaps (CDSs)}
In a CDS in-line with the concept outlined in Section 2.7 of [45], counterparties exchange a stream of coupon payments for a single default protection payment in the event of a default by a reference entity. The coupon payments have to be paid by the one counterparty either until maturity or, if a contractually agreed upon default event occurs earlier, only up to but including that incident. The other counterparty is obliged to pay a contingent default compensation in the case of the predefined event. Otherwise, if nothing the like happens until maturity, no payments become due for the other counterparty. Correspondingly, the CDS spread $\cds_n(0,T)$ is defined implicitly as solution to
\begin{align*}
&\cds_n(0,T)\sum_{i=1}^n(T_i-T_{i-1})E_\Q\bigg[e^{-\integ{0}{T_i}{r_u}{u}}\i_{\{\tau\geq T_i\}}\bigg]\\
\ &\stackrel{!}{=}\sum_{i=1}^nE_\Q\bigg[e^{-\integ{0}{T_i}{r_u}{u}}(1-S_{T_i})\i_{\{\tau=T_i\}}\bigg],
\intertext{where $\tau:=\min\set{T_1,T_2,\hdots,T_n}{S_{T_i}<1}$. Under the suitable assumptions that we made, this equation simplifies to}
&\cds_n(0,T)\sum_{i=1}^n(T_i-T_{i-1})P(0,T_i)E_\Q\big[\i_{\{\tau\geq T_i\}}\big]\\
\ &\stackrel{!}{=}\sum_{i=1}^nP(0,T_i)\Big(E_\Q\big[\i_{\{\tau=T_i\}}\big]-E_\Q\big[S_{T_i}\i_{\{\tau=T_i\}}\big]\Big).
\end{align*}
Furthermore, as derived in the next two sections, CDS spreads can be calculated explicitly.

\subsection*{Probability of Default (PD)}
We proceed consecutively. Since $T_1$ is the first possible default time at all, we have $\pd(0,T_1):=E_\Q\big[\i_{\{\tau\leq T_1\}}\big]=E_\Q\big[\i_{\{\tau= T_1\}}\big]=1-E_\Q\big[\i_{\{\tau> T_1\}}\big]$ and $E_\Q\big[\i_{\{\tau\geq T_1\}}\big]=1$. As seen from time $T_0$, no default occurs at time $T_1$ with probability
\begin{align}E_\Q\big[\i_{\{\tau> T_1\}}\big]&=E_\Q\big[\i_{\{X_{T_1}=0\}}\big]=\lim_{u\to-\infty}E_\Q\big[e^{uX_{T_1}}\big]=\lim_{u\to-\infty}e^{\phi_X(T_1,u)+x_0\psi_X(T_1,u)}.\nonumber
\intertext{Provided that $\mu_X=0$, we get}
E_\Q\big[\i_{\{\tau> T_1\}}\big]&=1-\pd(0,T_1)=\bigg(\frac{1}{2}\lambda_X{\sigma_X}^2T_1+1\bigg)^{-\frac{2m_X}{\lambda_X{\sigma_X}^2}}e^{-\frac{2x_0}{{\sigma_X}^2T_1}}.\nonumber
\intertext{For $i\geq 2$, it holds $E_\Q\big[\i_{\{\tau= T_i\}}\big]=\pd(0,T_i)-\pd(0,T_{i-1})$ and $E_\Q\big[\i_{\{\tau\geq T_i\}}\big]=1-\pd(0,T_{i-1})$. Similarly, we get}
E_\Q\big[\i_{\{\tau> T_2\}}\big]&=E_\Q\Big[\i_{\{X_{T_1}=0\}}E_\Q\big[\i_{\{X_{T_2}=0\}}\big|\M{F}_{T_1}\big]\Big]\nonumber\\
&=E_Q\bigg[\i_{\{X_{T_1}=0\}}\lim_{u\to-\infty}e^{\phi_X(T_2-T_1,u)+X_{T_1}\psi_X(T_2-T_1,u)}\bigg].\nonumber
\intertext{Provided that $\mu_X=0$, induction yields the formula}
E_\Q\big[\i_{\{\tau> T_i\}}\big]&=1-\pd(0,T_i)=e^{-\frac{2x_0}{{\sigma_X}^2T_1}}\prod_{j=1}^i\bigg(\frac{1}{2}\lambda_X{\sigma_X}^2(T_j-T_{j-1})+1\bigg)^{-\frac{2m_X}{\lambda_X{\sigma_X}^2}}.\nonumber
\end{align}

\subsection*{Loss Given Default (LGD)}
 Again, we proceed consecutively. As seen from time $T_0$, recovery given default at time $T_1$ is given by
\begin{align}
E_\Q\big[S_{T_1}\i_{\{\tau=T_1\}}\big]&=E_\Q\big[S_{T_1}\big]-E_\Q\big[S_{T_1}\i_{\{S_{T_1}=1\}}\big]\nonumber\\
&=F(0,T_1)-E_\Q\big[\i_{\{S_{T_1}=1\}}\big]\nonumber\\
&=F(0,T_1)-\big(1-\pd(0,T_1)\big).\nonumber
\intertext{Analogously,}
E_\Q\big[S_{T_2}\i_{\{\tau=T_2\}}\big]&=E_\Q\Big[\i_{\{S_{T_1}=1\}}E_\Q\big[S_{T_2}\i_{\{S_{T_2}<1\}}\big|\M{F}_{T_1}\big]\Big]\nonumber\\
&=E_\Q\bigg[\i_{\{S_{T_1}=1\}}\Big(F(T_1,T_2)-E_\Q\big[\i_{\{S_{T_2}=1\}}\big|\M{F}_{T_1}\big]\Big)\bigg]\nonumber\\
&=e^{\phi_X(T_2-T_1,-1)}E_\Q\big[\i_{\{\tau>T_1\}}\big]-E_\Q\big[\i_{\{\tau>T_2\}}\big].\nonumber
\intertext{Induction yields for $i\geq 2$ the formula}
E_\Q\big[S_{T_i}\i_{\{\tau=T_i\}}\big]&=e^{\phi_X(T_i-T_{i-1},-1)}E_\Q\big[\i_{\{\tau>T_{i-1}\}}\big]-E_\Q\big[\i_{\{\tau>T_i\}}\big]\nonumber\\
&=e^{\phi_X(T_i-T_{i-1},-1)}\big(1-\pd(0,T_{i-1})\big)-\big(1-\pd(0,T_i)\big).\label{eq:rgd}
\end{align}
The formula~\eqref{eq:rgd} is also valid for $i=1$ if it holds $x_0=0$ (i.e., $S_0=1$) and if we adhere to the convention $\pd(0,T_0):=0$.\\

\subsection*{CDS Spreads}
All in all, provided that $x_0=0$, we end up with the explicit formula
\begin{equation}\label{eq:cds}\cds_n(0,T)=\frac{\sum_{i=1}^nP(0,T_i)\big(1-e^{\phi_X(T_i-T_{i-1},-1)}\big)\big(1-\pd(0,T_{i-1})\big)}{\sum_{i=1}^n(T_i-T_{i-1})P(0,T_i)\big(1-\pd(0,T_{i-1})\big)}.\end{equation}
In this case, the factor $e^{\phi_X(T_i-T_{i-1},-1)}$ coincides with $F(0,T_i-T_{i-1})$; see \eqref{eq:f}. If we had $x_0>0$, the expression $e^{\phi_X(T_i-T_{i-1},-1)}$ in the numerator of \eqref{eq:cds} for $i=1$ (and only for $i=1$) would need to be replaced by $F(0,T_1)$. In the special case $n=1$, the formula reduces to
\[\cds_1(0,T)=\frac{1-F(0,T)}{T}.\]
If $x_0=0$ and the partition is chosen equidistant, then it holds
\begin{equation}\label{eq:cdslight}\cds_n(0,T)=\frac{1-F(0,T/n)}{T/n}.\end{equation}

\subsection*{The Calibration Task}

We parameterised the two term structures $T\longmapsto P(0,T)$ and $T\longmapsto\tilde{P}(0,T)$ in terms of a deterministic function $\varphi$ and eight model parameters $r_0$, $b_x$, $\beta_x$, $\sigma_x$, $\lambda_X$, $m_X$, $\sigma_X$ as well as $x_0$; the recovery rate is not assumed to be self-exciting (i.e., $\mu_X=0$) and starts at $S_0=1$. Given that we also know the term structure $T\longmapsto\mathbb{L}(0,T)$, the mapping from these eight parameters onto market quotes $V_{n,c}(0,T)$, $\tilde{V}_{n,c}(0,T)$ and $\cds_n(0,T)$ is straightforward. The other way round, however, is non-trivial. The bootstrapping of $P(0,T)$, $\tilde{P}(0,T)$ and $\mathbb{L}(0,T)$ from marketable products is often cumbersome. We propose the following non-parametric calibration procedure for the initial yield curves on an equidistant partition of $[0,T]$:
\begin{enumerate}
\item[(i)] Bootstrap the non-defaultable term structure $T\longmapsto P(0,T)$ from liquid government bonds utilising equation~\eqref{eq:govbonds}. This can be achieved in the sense of least squares by solving the corresponding normal equation.
\item[(ii)] Derive the defaultable term structure $T\longmapsto\tilde{P}(0,T)$ from CDS spreads utilising the relation $\tilde{P}(0,T)=F(0,T)P(0,T)$ and equation~\eqref{eq:cdslight}.
\item[(iii)] Bootstrap the illiquidity premium $T\longmapsto\mathbb{L}(0,T)$ from issued corporate bonds according to \eqref{eq:corpbonds} in order to explain the residual spread of corporate bonds above their governmental counterparts. To this end, one can proceed analogously as in (i).
\end{enumerate}
This is a simple yet powerful algorithm to take a snapshot of the current market situation. For dynamic approaches, liquid derivatives should be incorporated in order to account for the volatility surface of the term structure.

\begin{remark}[Machine Learning]
If we parameterise $\varphi$ in terms of $\varphi(t)=f_1\varphi_1(t)+f_2\varphi_2(t)+f_3\varphi_3(t)$, where $f_k\in\R$ and $t\longmapsto\varphi_k(t)$ is a suitably chosen set of basis functions/principal components for $k=1,2,3$, then the inverse of the mapping from the parameter set $\big\{r_0,b_x,\beta_x,\sigma_x,\lambda_X,m_X,\sigma_X,f_1,f_2,f_3\big\}$ onto the backed-out grid of the quotes $\big\{P(0,T_i),\tilde{P}(0,T_i),\mathbb{L}(0,T_i)\,\big|\,T_i=iT/n\textrm{ for }i=0,1,2,\hdots,n\big\}$ can be learnt by a sufficiently complex neural network. The target function in the minimisation can certainly be extended to financial derivatives with optionalities.
\end{remark}

\bibliographystyle{amsplain}

\end{document}